# Martensitic Transformation in Crystal-Amorphous Superlattices of NiTi Shape Memory Alloy

Bhavna Singh, Shivam Tripathi*

Department of Materials Science and Engineering, Indian Institute of Technology Kanpur, Kanpur, Uttar Pradesh, India 208016

*Corresponding author: Shivam Tripathi (shivamt@iitk.ac.in)

## Abstract

Shape memory alloys (SMAs) exhibit unique thermo-mechanical properties arising from reversible martensitic transformation. Tailoring these properties through coherent second phases is effective but is often limited by the difficulty of identifying thermodynamically compatible phases. Crystal-amorphous superlattices (CAS), in which the second phase is derived from the base material itself, provide an attractive alternative. However, the influence of partial amorphization on martensitic transformation and thermo-mechanical behavior remains largely unexplored. Here, large-scale molecular dynamics simulations are used to investigate temperature- and stress-induced martensitic transformations in CAS-NiTi with different crystalline phase fractions. Partial amorphization fundamentally modifies the transformation pathway, introducing an initial continuous (second-order-like) transformation before the conventional first-order martensitic transformation. The amorphous phase also enhances transformation reversibility, reducing the thermal hysteresis from 275 K in fully crystalline NiTi to 95-110 K in CAS-NiTi. Simultaneously, the elastic modulus and the critical stress for stress-induced martensitic transformation increase by approximately 60–90%, depending on the crystalline fraction. These improvements originate from heterogeneous nucleation at crystal-amorphous interfaces, retained austenite that promotes the reverse transformation, and mechanical constraint imposed by the amorphous phase. These findings establish crystal-amorphous superlattices as a promising microstructural design strategy for tailoring the thermo-mechanical performance of shape memory alloys.

# 1. Introduction

SMAs belong to a special class of materials that undergo solid-solid reversible martensitic transformation between a high-temperature, high-symmetry austenite phase and a low-temperature, low-symmetry martensite phase. This transformation underpins several interesting thermo-mechanical characteristics including shape memory behavior, superelasticity, and pseudo-plasticity. [1-3] These characteristics make them useful materials for various applications in automotive [4-6], aerospace [7-9], construction [9-11], micro-electro-mechanical systems (MEMS) [12-13], biomedical [9, 13-15] and others.

The number of naturally occurring SMAs is limited, and each comes with an inherent, fixed set of properties. Therefore, to design SMAs for specific applications, tunability is needed. Traditionally, tuning is achieved through alloying additions [16-18] and processing conditions [18-20] that control the microstructure. More recently, the incorporation of a coherent second phase within the base martensitic matrix via precipitation or superlattices based structures, has been shown to enhance the thermo-mechanical properties in an extraordinary manner [21-28].

For example, Reeve et al. [21] demonstrated using MD simulations that incorporating a NiAl second phase into the base martensitic $Ni_{63}Al_{37}$, in either a nanowire or nanolaminate configuration, can produce ultra-low stiffness in heteroepitaxial metamaterials. Tripathi et al. [22] used first-principles calculations to study martensitic transformations in Mg-Sc, and found that integrating pure Mg as a coherent second phase raises the transition temperature from -100°C in bulk Mg-Sc to near room temperature, considerably expanding the application space for Mg-based SMAs. They also showed martensitic transformation can be induced in a composite of two otherwise non-transforming materials, as in MgLi/Mg superlattices [23]. Experimentally, coherent $Ti_2Cu$ precipitates in a Ni-Ti-Cu alloy were shown to give no measurable fatigue after 10 million actuation cycles [25], and a similar effect was seen with nanoscale $Ni_{50}Al_{50}$ precipitates in NiAl-based SMAs via MD simulations [26]. Disordered omega precipitates in Ni-Co-Fe-Ga SMAs have also been shown to enable a novel non-hysteretic superelastic response [27].

Separately, Hua et al. [28] characterized a Ni-Ti crystalline-amorphous nanocomposite and reported high strength along with good fatigue resistance. This result is notable because identifying a suitable coherent second phase is inherently difficult [29], and their work suggests an alternative: rather than introducing a different crystalline phase, amorphizing part of the same base matrix can itself serve as the coherent second phase. However, the study did not examine the role of the amorphous phase in the thermally-induced martensitic transformations within these systems. The atomic-scale mechanisms responsible for these behaviors, as well as the influence of varying the relative fractions of crystalline and amorphous phases on both thermally and mechanically induced martensitic transformations, remain unexplored.

Addressing this gap, our study characterizes the thermal and stress-induced martensitic transformation in CAS-NiTi with varying crystalline fractions (100%, 66%, 53%, and 43% crystalline fractions). We find that amorphizing part of the matrix substantially improves the thermo-mechanical properties, with a significant reduction in thermal hysteresis and increases in both Young's modulus and the critical stress for superelasticity.

# 2. Simulation Details

MD simulations were performed using the Large-scale Atomic/Molecular Massively Parallel Simulator (LAMMPS) [30]. Atomic structures were visualized using the Open Visualization Tool (OVITO) [31]. We used the Parrinello-Rahman barostat [32] and the Nose-Hoover thermostat [33-34], with a time step of 1 fs for all of our simulations. Periodic boundary conditions were used in all three dimensions. We used the second-nearest neighbor modified embedded atom method (2NN MEAM) interatomic potential developed for characterizing martensitic phase transformation in NiTi SMAs [35]. This potential has been shown to reliably reproduce the martensitic transformation behavior of NiTi and has been widely used in previous molecular dynamics studies [36-41]. The procedures used to construct CAS-NiTi structures and to characterize thermally and mechanically induced transformations are described in Sections 2.1–2.3.

## 2.1. Crystal-amorphous superlattices

We started with a NiTi B2 supercell containing 1.0240 million atoms, with dimensions of 24 nm, 24 nm, and 24 nm along the [100], [010], and [001] directions, respectively. This fully crystalline structure is hereafter referred to as the 100% Xtal (crystalline phase fraction) configuration. With the initial aim of generating CAS-NiTi with 25% Xtal, 50% Xtal, and 75% Xtal, the simulation box was partitioned along the [001] direction into crystalline and amorphous regions whose number of atoms corresponded to the target crystalline fractions (25%, 50%, and 75%). For all three CAS-NiTi (25% Xtal, 50% Xtal, and 75% Xtal), the atoms designated for the crystalline phase were maintained at 500 K, while those intended to form the amorphous regions subjected to the following thermal cycle: (i) heating from 500 K to 5000 K at a rate of 5 K/ps, (ii) annealing at 5000 K for 1000 ps, and (iii) cooling from 5000 K back to 500 K at the same rate. The entire simulation cell was subsequently relaxed in the isothermal-isobaric (NPT) ensemble at 500 K for 1000 ps. SI Figure 1 shows the representative temperature versus time profile used to construct CAS-NiTi structures.

Figure 1 shows the snapshots of intended 75% Xtal, 50% Xtal, and 25% Xtal CAS-NiTi structures after amorphization and NPT relaxation at 500 K. Atoms are color-coded according to Polyhedral Template Matching (PTM) analysis [42] implemented in OVITO. PTM identifies the local crystal structure by comparing atomic neighbourhoods with ideal reference lattices using the root mean square deviation (RMSD). Atoms identified as BCC are classified as austenite, HCP or FCC atoms as martensite, while unidentified atoms are treated as amorphous. The crystalline fractions obtained after amorphization and subsequent NPT relaxation differ from the intended target values, yielding approximately 43%, 53%, and 66% crystalline material instead of the designed 25%, 50%, and 75%, respectively. This difference originates from atomic diffusion and structural reconstruction near the crystal-amorphous interfaces during the high-temperature amorphization process.

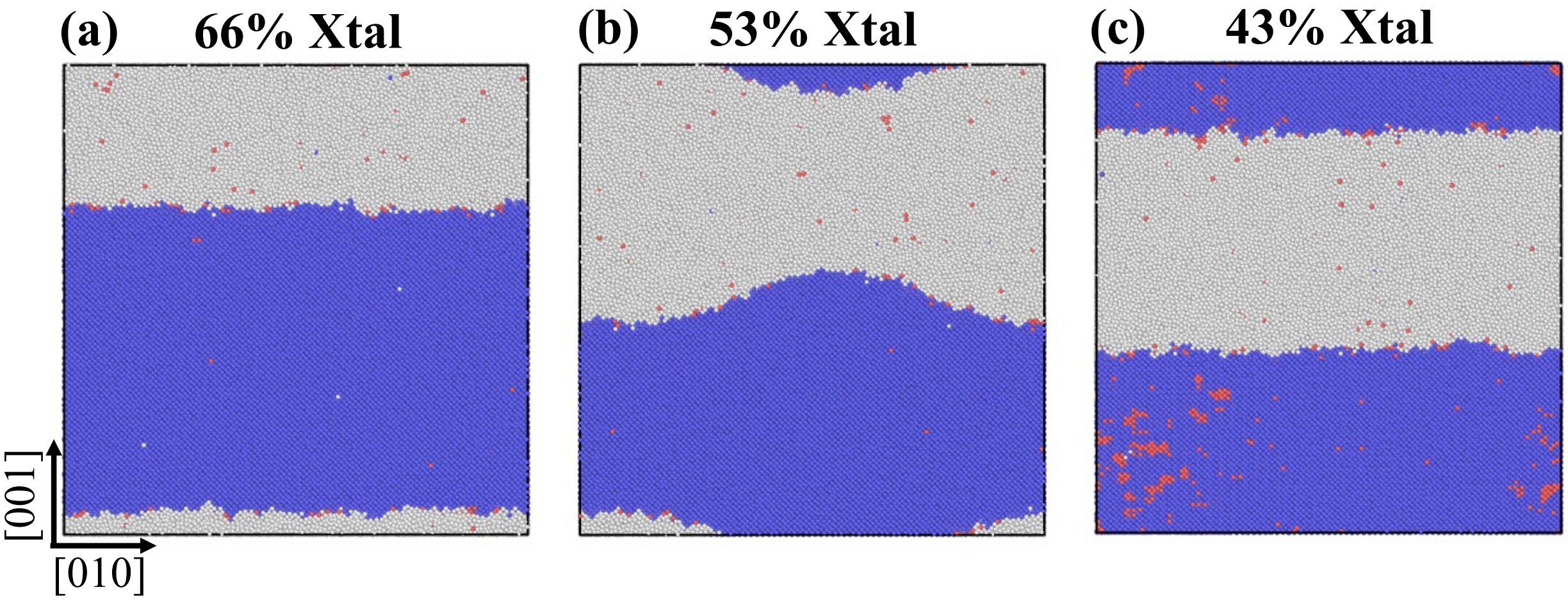

**Figure 1.** Crystal–amorphous superlattice (CAS-NiTi) structures after amorphization and subsequent NPT relaxation at 500 K for (a) 66% Xtal, (b) 53% Xtal, and (c) 43% Xtal systems. Atoms are color-coded according to the Polyhedral Template Matching (PTM) analysis: austenite (blue), amorphous (grey), and martensite (red).

## 2.2. Temperature-induced martensitic transformation

To investigate the temperature-induced martensitic transformation, we performed cooling-heating simulations on the 43% Xtal, 53% Xtal, 66% Xtal, and 100% Xtal CAS-NiTi systems. Starting from the NPT-relaxed structures at 500 K, each system was cooled from 500 K to 10 K at a rate of 5 K/ps. The resulting structures were subsequently reheated from 10 K to 550 K at the same rate. The thermal hysteresis was calculated as a difference between the austenite finish temperature ($A_f$), and the martensite start temperature ($M_s$). The transformation temperatures were determined from the variation of the martensite fraction with temperature.

## 2.3. Stress-induced martensitic transformation

Stress-induced martensitic transformation was studied using uniaxial loading and unloading simulations. Simulations were first performed at 500 K along two loading directions: [001], perpendicular to the crystal-amorphous interface, and [100], parallel to the interface. Because the [100] and [010] directions are crystallographically equivalent and both are parallel to the interface, only the [100] direction was considered.

Starting from an NPT-relaxed CAS-NiTi structures, we performed loading simulations by elongating the simulation cell along the tensile axis [100] and [001] up to 12% strain, using a strain rate of $5x10^7$ $s^{-1}$. The lateral dimensions and angles of the simulation cell were allowed to evolve to maintain a target pressure of 1 atm. Unloading simulations were initiated from the 8% and 12% strained configurations under stress-controlled conditions by gradually reducing the applied stress along the tensile axis to 1 atm over 1600 ps and 2400 ps, respectively, matching the loading durations. We investigated stress-induced transformations for all our systems: 43% Xtal, 53% Xtal, 66% Xtal, and 100% Xtal CAS-NiTi. Mechanical hysteresis was quantified as the area enclosed by the loading and unloading stress–strain curves.

The elastic modulus was obtained from the slope of the linear elastic region of the loading stress-strain curve. Linear fits over strain intervals of 0-0.01, 0-0.02, and 0-0.03 yielded nearly identical values (within a few GPa). Therefore, the 0-0.03 strain interval was used throughout this work. The critical stress for martensitic transformation was determined using the 0.2% offset method (analogous to the conventional 0.2% yield stress), together with the maximum stress marking the onset of the stress plateau during loading.

# 3. Results and Discussion

## 3.1. Temperature-induced martensitic transformation

We performed heating and cooling simulations to study thermally induced martensitic transformation. Figure 2 shows the evolution of the martensitic phase fraction as a function of temperature during these simulations. The NPT-relaxed structures at 500 K are predominantly in the austenite phase within the crystalline regions of all CAS-NiTi systems. Upon cooling, the fully crystalline (100% Xtal) system undergoes an abrupt first-order martensitic transformation beginning at the $M_s$. However, CAS-NiTi systems with finite amorphous content, particularly the 43% Xtal and 53% Xtal systems, exhibit an initial uncharacteristic, continuous (second-order-like) martensitic transformation before the onset of the conventional first-order transformation. The coexistence of a gradual increase in martensite fraction followed by an abrupt first-order transformation suggests that the crystal-amorphous interfaces facilitate localized martensitic nucleation prior to the collective transformation of the remaining crystalline regions. Therefore, $M_s$ is defined as the onset of the first-order transformation. With further cooling, the martensite fraction increases until it saturates. During the heating phase, the martensite transforms back to the austenite phase, completing the reverse transformation at the $A_f$. Consistent with its first-order nature, the transformation is accompanied by an abrupt change in the simulation cell volume, as shown in SI Figure 2.

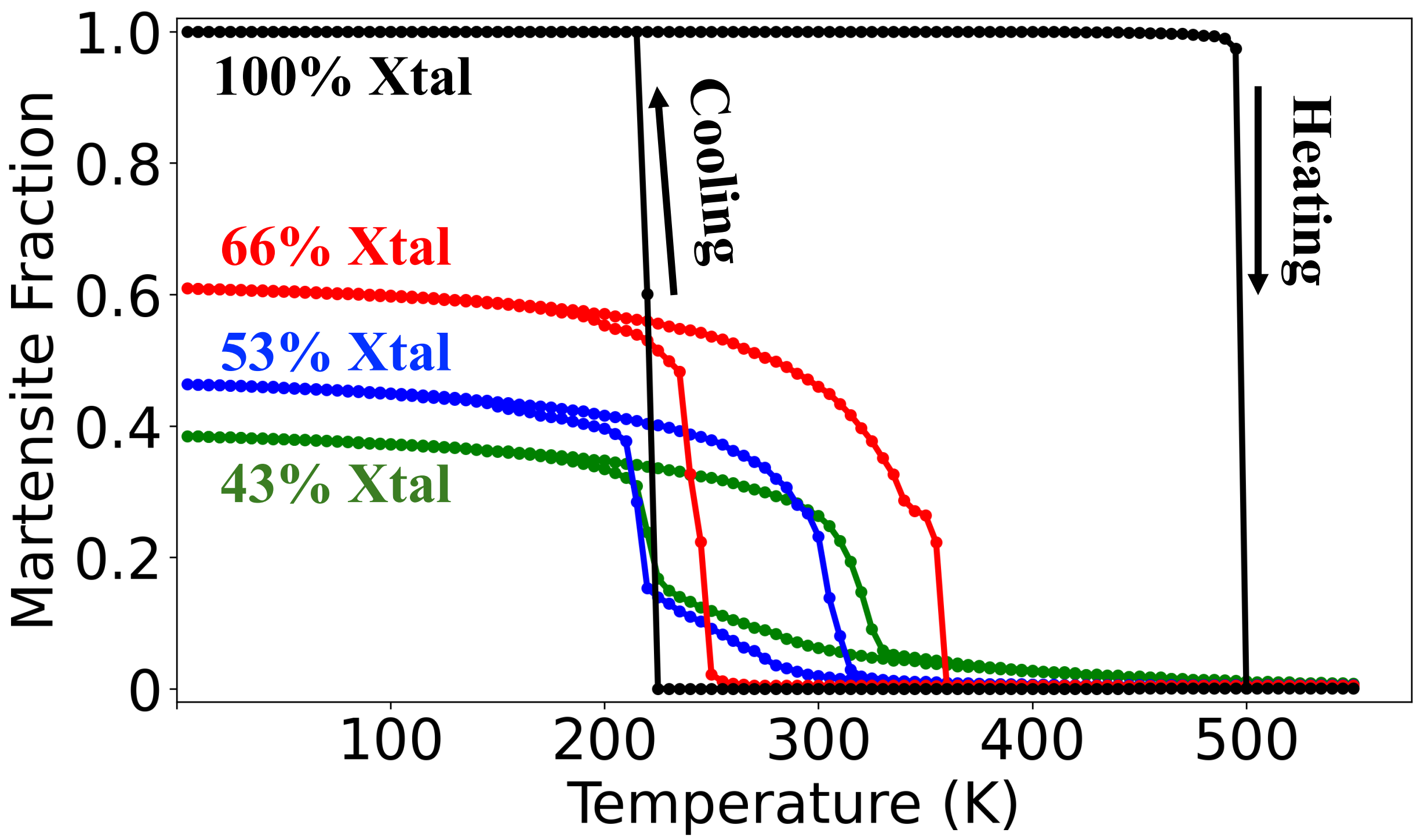

**Figure 2.** Martensite fraction as a function of temperature during heating and cooling simulations for various CAS-NiTi. Green, blue, red, and black colors are used for 43%, 53%, 66%, and 100% Xtal CAS-NiTi.

The introduction of an amorphous phase has little influence on $M_s$ but significantly lowers $A_f$, indicating enhanced stabilization of the austenite phase. As a result, the thermal hysteresis ($A_f$ - $M_s$) is substantially reduced. The values of $M_s$, $A_f$, and thermal hysteresis values for various CAS-NiTi systems are reported in Table 1. Introducing approximately 34% amorphous phase reduces the thermal hysteresis by nearly 60%, whereas further increases in amorphous content produce only marginal additional reductions. We note that the thermal hysteresis predicted for fully crystalline NiTi is larger than the values typically reported experimentally [43]. This difference can be attributed to two primary factors. First, the present simulations consider defect-free single-crystal NiTi, whereas experimental measurements are generally performed on polycrystalline materials containing grain boundaries, dislocations, and other heterogeneous nucleation sites. The absence of such defects increases the barrier for both forward and reverse transformations, leading to lower predicted $M_s$ and higher $A_f$, and consequently a larger thermal hysteresis. Second, the heating and cooling rates used in MD simulations are several orders of magnitude higher than those accessible experimentally, which further shifts the transformation temperatures away from equilibrium. Nevertheless, the predicted transformation behavior is consistent with previous molecular dynamics studies of NiTi with the same interatomic potential [41]. Since all CAS-NiTi systems were simulated using an identical computational framework, the relative changes in transformation temperatures and thermal hysteresis arising from the introduction of the amorphous phase remain robust.

**Table 1.** Martensitic start temperature ($M_s$), austenite finish temperature ($A_f$), and thermal hysteresis ($A_f$ - $M_s$) for the crystal-amorphous superlattice (CAS-NiTi) systems. The transformation temperatures were determined from the temperature dependence of the martensite fraction during the cooling-heating simulations shown in Figure 1.

| **CAS-NiTi System** | **$M_s$ (K)** | **$A_f$ (K)** | **Thermal Hysteresis (K)** |
|---|---|---|---|
| 100 % Xtal | 225 ± 5 | 500 ± 5 | 275 ± 10 |
| 66 % Xtal | 250 ± 5 | 360 ± 5 | 110 ± 10 |
| 53 % Xtal | 220 ± 5 | 315 ± 5 | 95 ± 10 |
| 43 % Xtal | 225 ± 5 | 330 ± 5 | 105 ± 10 |

To gain atomistic insight, Figure 3 (100% Xtal and 53% Xtal CAS-NiTi) and SI Figure 3 (66% Xtal and 43% Xtal CAS-NiTi) shows atomic snapshots at various stages of the forward and reverse martensitic transformations. The snapshots correspond to the following temperatures: during cooling at 500 K, $M_s$ - 5 K, $M_s$, and 10 K; and during heating at 10 K, $A_f$ - 5 K, $A_f$ , and 550 K.

The fully crystalline system develops a single dominant martensite variant during cooling (see Figure 3), In contrast, the CAS-NiTi systems develop multiple martensite variants (see Figure 3 and SI Figure 3), indicating simultaneous nucleation at multiple sites. Additionally, we observe nearly complete martensitic transformation in the fully crystalline system. However, a fraction of the austenite phase remains untransformed in the CAS-NiTi systems containing an amorphous phase. These retained austenite regions provide pre-existing nuclei for the reverse transformation during heating, thereby facilitating austenite growth, lowering the $A_f$, and consequently reducing the thermal hysteresis.

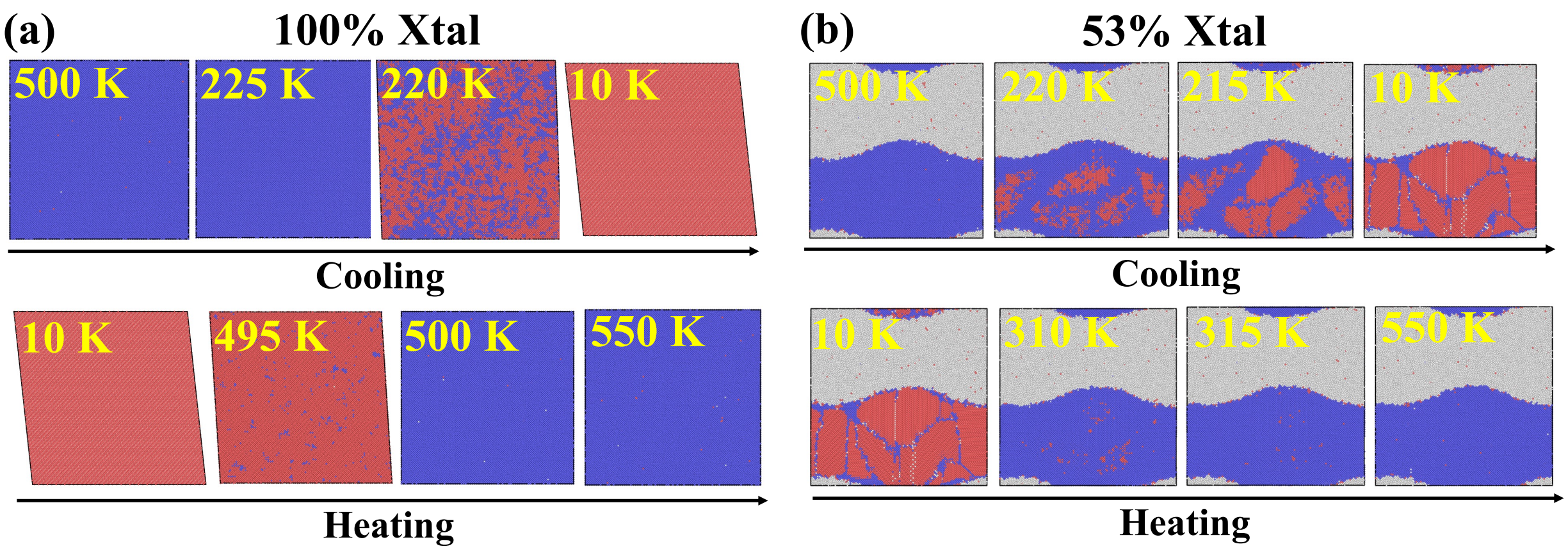

**Figure 3.** Atomic configurations during the forward (cooling) and reverse (heating) martensitic transformations for (a) 100% Xtal and (b) 53% Xtal CAS-NiTi systems. Snapshots correspond to selected temperatures during cooling (500 K, $M_s$, $M_s - 5$K, and 10 K) and heating (10 K, $A_f - 5$K, $A_f$, and 550 K). Atoms are colored according to the Polyhedral Template Matching (PTM) analysis: austenite (blue), martensite (red), and amorphous (gray).

Overall, these results demonstrate that the CAS-NiTi composites not only reduce the thermal hysteresis but also fundamentally modifies the martensitic transformation pathway by introducing an initial continuous transformation prior to the conventional first-order transformation, promoting heterogeneous nucleation, generating multiple martensite variants, and retaining austenite that facilitates the reverse transformation.

## 3.2. Stress-induced martensitic transformation

The stress-induced martensitic transformation and superelastic response were examined through loading-unloading simulations at 500 K (see Section 2.2 for details). Figure 4 presents the stress-strain response of all CAS-NiTi systems at 500 K under uniaxial loading along the [001] (perpendicular to the crystal-amorphous interface) and [100] (parallel to the interface) directions. The response is similar for both loading directions, indicating that the loading orientation has only a minor effect on the superelastic behavior. Therefore, the remainder of this study focuses on the [001] loading direction. Two unloading cases are considered: one from a maximum strain of 8%, and another from 12%. As anticipated, the loading and unloading curves for the 100% Xtal CAS-NiTi nearly coincide, indicating negligible hysteresis. This behavior is attributed to the absence of defects in the single-crystal system, which limits mechanisms for mechanical energy dissipation. In contrast, the 43%, 53%, and 66% Xtal CAS-NiTi systems exhibit characteristic superelastic behavior accompanied by increasing mechanical hysteresis and remnant strain as the amorphous phase fraction increases, see Figure 5. This behavior indicates that the amorphous regions and crystal-amorphous interfaces promote mechanical energy dissipation while simultaneously accommodating irreversible deformation.

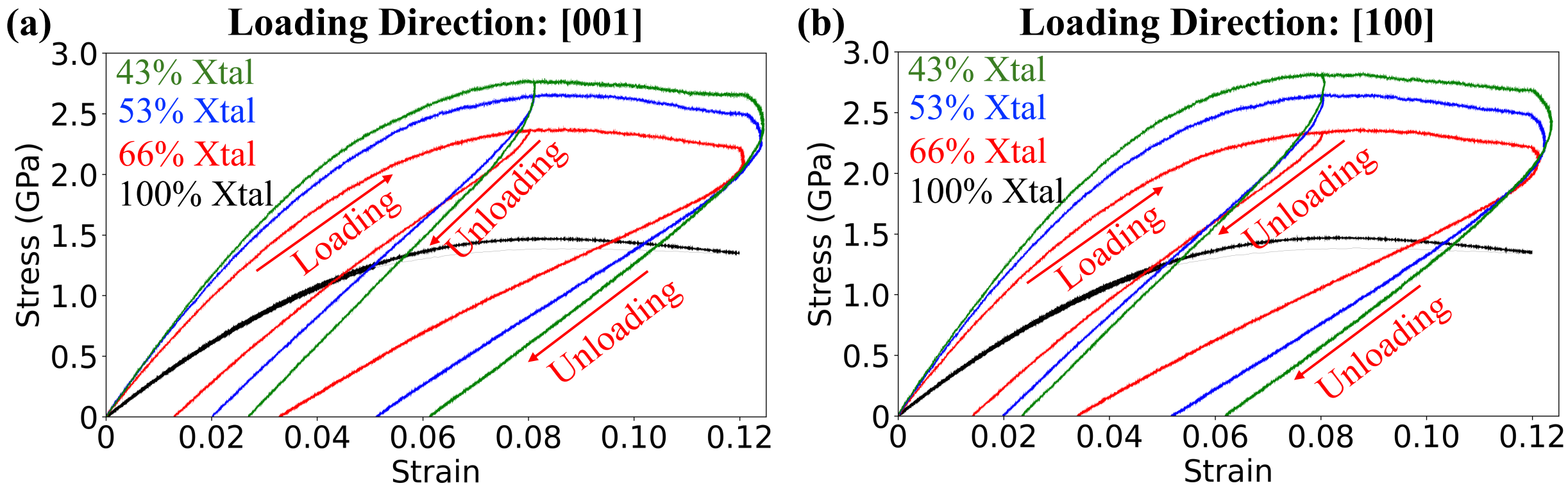

**Figure 4.** Stress-strain response during stress-induced martensitic transformation in 100% Xtal, 66% Xtal, 53% Xtal, and 43% Xtal CAS-NiTi systems at 500 K. Uniaxial loading was applied along (a) the [001] direction, perpendicular to the crystal-amorphous interface, and (b) the [100] direction, parallel to the interface. In each case, unloading was performed from maximum strains of 8% and 12%.

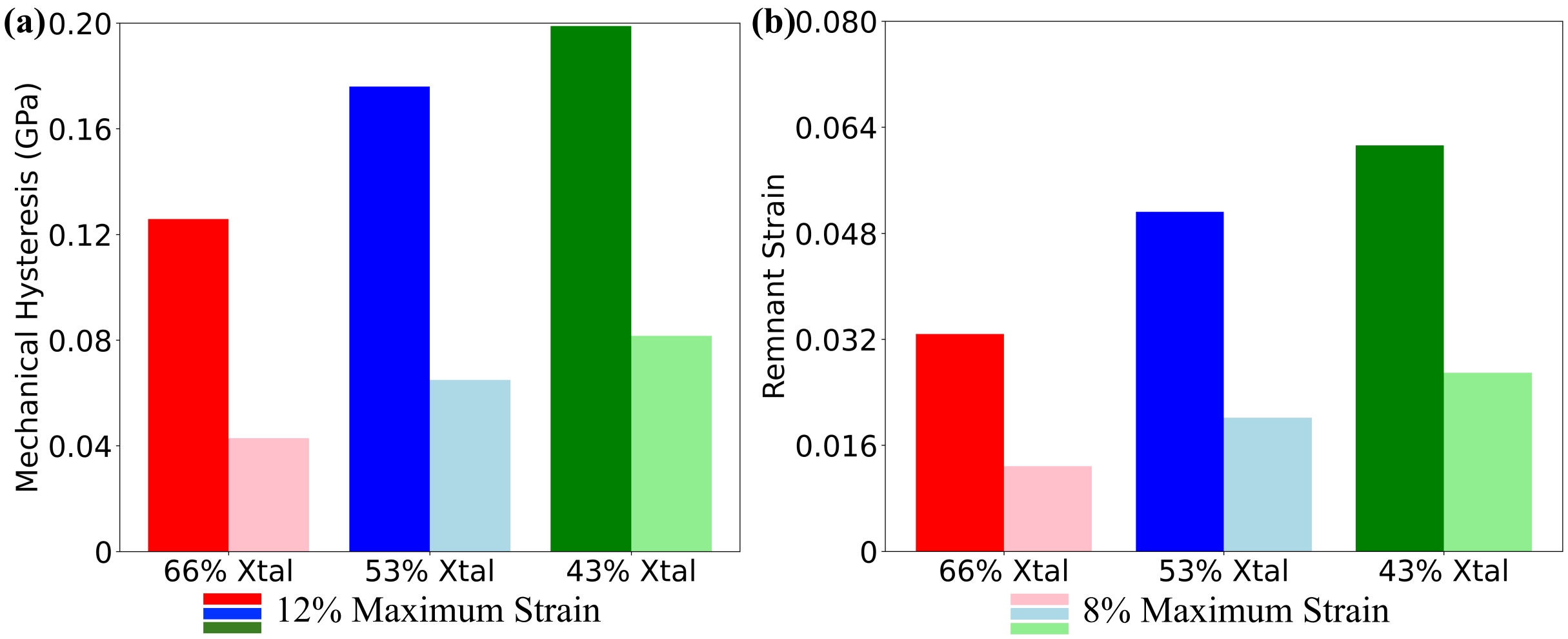

**Figure 5.** Comparison of (a) mechanical hysteresis (area under stress-strain curve) and (b) remnant strain for the investigated CAS-NiTi systems following unloading from maximum strains of 12% and 8% under uniaxial tensile loading along the [001] direction at 500 K

The incorporation of the amorphous phase substantially increases the elastic modulus, from 29 ± 1 GPa for fully crystalline NiTi to 46 ± 1, 52 ± 1, and 55 ± 1 GPa for the 66%, 53%, and 43% Xtal CAS-NiTi systems, respectively. Likewise, both the 0.2% offset critical stress and the maximum pre-plateau stress increase systematically with increasing amorphous content (Table 2). The procedures used to determine these quantities from the loading stress-strain response are described in Section 2.2 and illustrated in SI Figure 4. These results indicate that the amorphous phase mechanically constrains stress-induced martensitic transformation, requiring higher applied stresses to initiate and propagate the transformation.

**Table 2.** Elastic modulus, 0.2% offset critical stress for stress-induced martensitic transformation, and maximum pre-plateau stress obtained from the loading stress-strain response of the investigated CAS-NiTi systems. The procedure used to determine these quantities is illustrated in SI Figure 4.

| **CAS-NiTi System** | **Modulus (GPa)** | **0.2% offset critical stress (GPa)** | **Maximum pre-plateau stress (GPa)** |
|---|---|---|---|
| 100 % Xtal | 29 ± 1 | 0.95 | 1.49 |
| 66 % Xtal | 46 ± 1 | 1.50 | 2.40 |

| 53 % Xtal | 52 ± 1 | 1.64 | 2.68 |
|---|---|---|---|
| 43 % Xtal | 55 ± 1 | 1.82 | 2.79 |

Snapshots at various stages of the loading (pre-loading, 3.5% strain (near the 0.2% offset critical stress), 8% strain (near the maximum pre-plateaus stress), and 12% strain) and unloading (pre-unloading and post-unloading configurations) simulations for all studied CAS-NiTi systems are shown in Figure 6 and SI Figure 5. CAS-NiTi systems containing amorphous phase exhibit minor martensitic regions in the crystalline phase even in the pre-deformed structures, particularly in the 43% Xtal system. These are likely induced by local stress concentrations associated with the internal stress fields generated at the crystal-amorphous interfaces. At 3.5% strain, initial martensite nucleation is observed. Following nucleation, the martensite fraction increases gradually up to 8% strain before rising more rapidly as the strain reaches 12%. Upon unloading from both the 12% and 8% strained configurations, reverse transformation occurs; however, a finite remnant martensite persists in the unloaded structures.

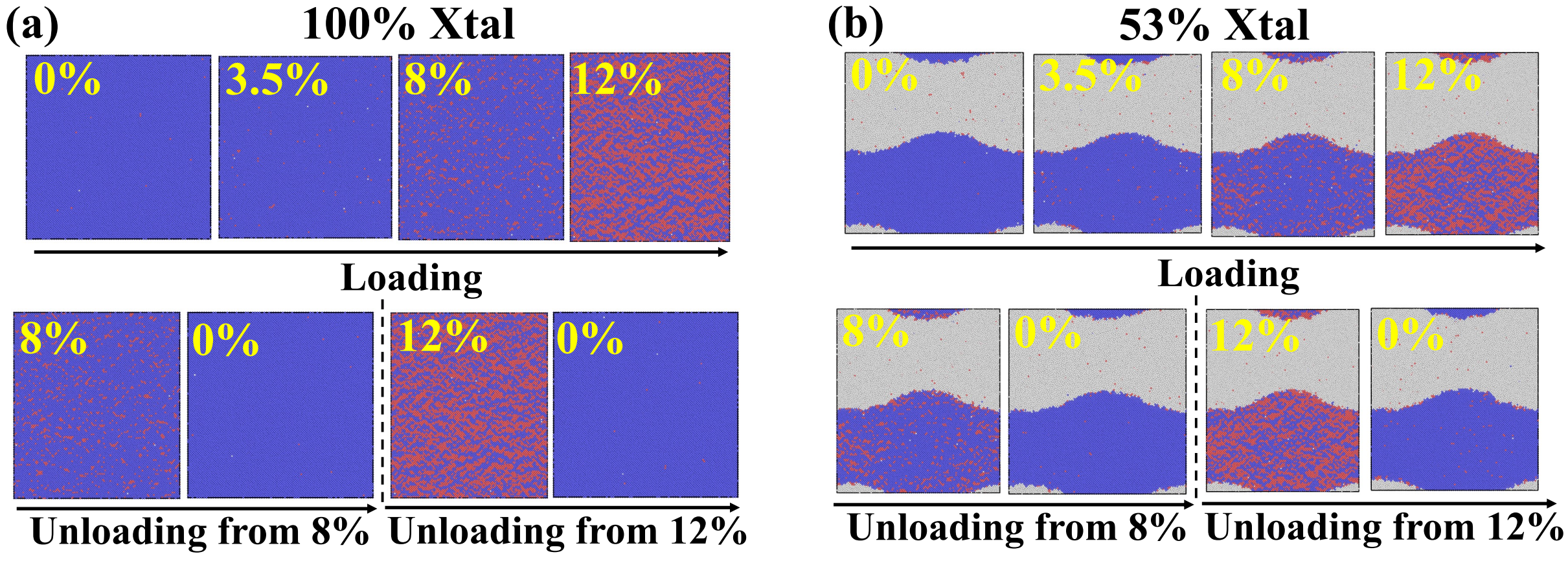

**Figure 6.** Atomic snapshots illustrating the evolution of stress-induced martensitic transformation during loading and unloading in (a) 100% Xtal and (b) 53% Xtal CAS-NiTi systems under uniaxial tensile loading along the [001] direction at 500 K. Snapshots during loading correspond to 0%, 3.5%, 8%, and 12% strain. Unloading is performed separately from maximum strains of 8% and 12%, and the corresponding unloaded configurations are shown to illustrate the reversibility of the transformation and the remnant martensite

Overall, these results demonstrate that the amorphous phase imposes a mechanical constraint on the crystalline regions, leading to increased elastic modulus and higher stresses for stress-induced martensitic transformation. At the same time, the amorphous regions enhance mechanical energy dissipation through superelastic hysteresis while introducing a finite remnant strain associated with irreversible deformation. These findings demonstrate that CAS-NiTi structures provide an effective strategy for simultaneously tuning the stiffness, transformation stress, hysteresis, and superelastic recovery of NiTi through microstructural design.

# 4. Conclusions

In this study, we investigated the temperature and stress-induced martensitic transformation in various crystal-amorphous superlattices. Our simulations reveal that the presence of amorphous phase fundamentally modifies the transformation pathway by introducing a two-stage transformation consisting of an initial continuous (second-order-like) transformation followed by the conventional first-order martensitic transformation. The amorphous phase stabilizes the austenite phase and reduces the thermal hysteresis by nearly 60%, with little additional dependence on amorphous content beyond the initial introduction of the amorphous phase. Crystal-amorphous superlattices simultaneously increases the elastic modulus and the critical stress for stress-induced martensitic transformation, demonstrating that the amorphous phase mechanically constrains the crystalline regions. Specifically, the elastic modulus increases by approximately 60%, 80%, and 90%, while the critical stress for stress-induced martensitic transformation increases by roughly 60%, 70%, and 90% for the 66%, 53%, and 43% Xtal CAS-NiTi systems, respectively, relative to fully crystalline NiTi. Atomistic analysis revealed that these improvements originate from heterogeneous nucleation at crystal-amorphous interfaces, the formation of multiple martensite variants, retained austenite that facilitates the reverse transformation, and the mechanical constraint imposed by the amorphous phase on the surrounding crystalline regions. In summary, we demonstrate that tailoring the second phase by partially amorphizing the base martensitic material can yield significantly improved shape memory alloys in terms of thermo-mechanical performance. Although demonstrated here for binary NiTi, the proposed approach provides a general microstructural design strategy that can be extended to other shape memory alloy systems for tailoring thermo-mechanical performance.

## Data Availability

The data supporting the findings of this study are available within the main article and its Supporting Information. Additional data related to this work are available from the corresponding author upon reasonable request.

## Acknowledgement

The author(s) thank the Anusandhan National Research Foundation (ANRF), India, for financial support through the Prime Minister Early Career Research Grant (PM ECRG) [ANRF/ECRG/2024/000523/ENS]. Computational resources and support provided by PARAM Sanganak under the National Supercomputing Mission, Government of India, at the Indian Institute of Technology Kanpur are gratefully acknowledged.

## Declaration of generative AI and AI-assisted technologies in the writing process

During the preparation of this work, the author(s) used ChatGPT in order to restructure some of the sentences. After using this tool/service, the author(s) reviewed and edited the content as needed and take(s) full responsibility for the content of the publication.

# Supporting Information

## Martensitic Transformation in Crystal-Amorphous Superlattices of NiTi Shape Memory Alloy

Bhavna Singh, Shivam Tripathi*

Department of Materials Science and Engineering, Indian Institute of Technology Kanpur, Kanpur, Uttar Pradesh, India 208016

*Corresponding author: Shivam Tripathi (shivamt@iitk.ac.in)

## 1. Temperature-induced martensitic transformation

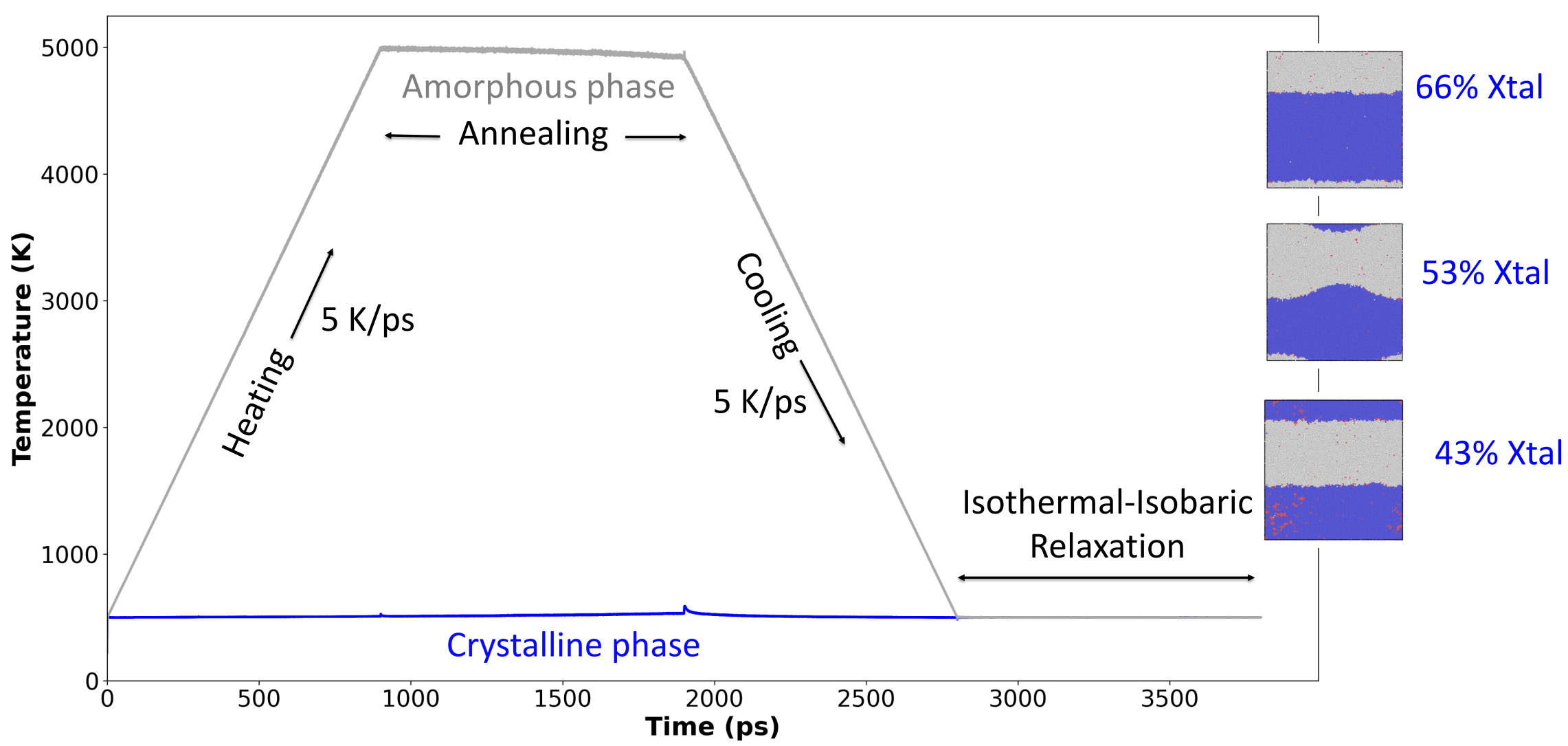

**SI Figure 1.** Temperature–time profile used to generate crystal–amorphous superlattice (CAS-NiTi) structures. Atoms designated to form the amorphous region were heated from 500 K to 5000 K at a rate of 5 K/ps, annealed at 5000 K, and subsequently cooled to 500 K at the same rate, while the crystalline region was maintained at 500 K throughout the process. The entire system was then relaxed under NPT conditions at 500 K. Representative CAS-NiTi structures with 66%, 53%, and 43% crystalline phase fractions after NPT relaxation are shown on the right.

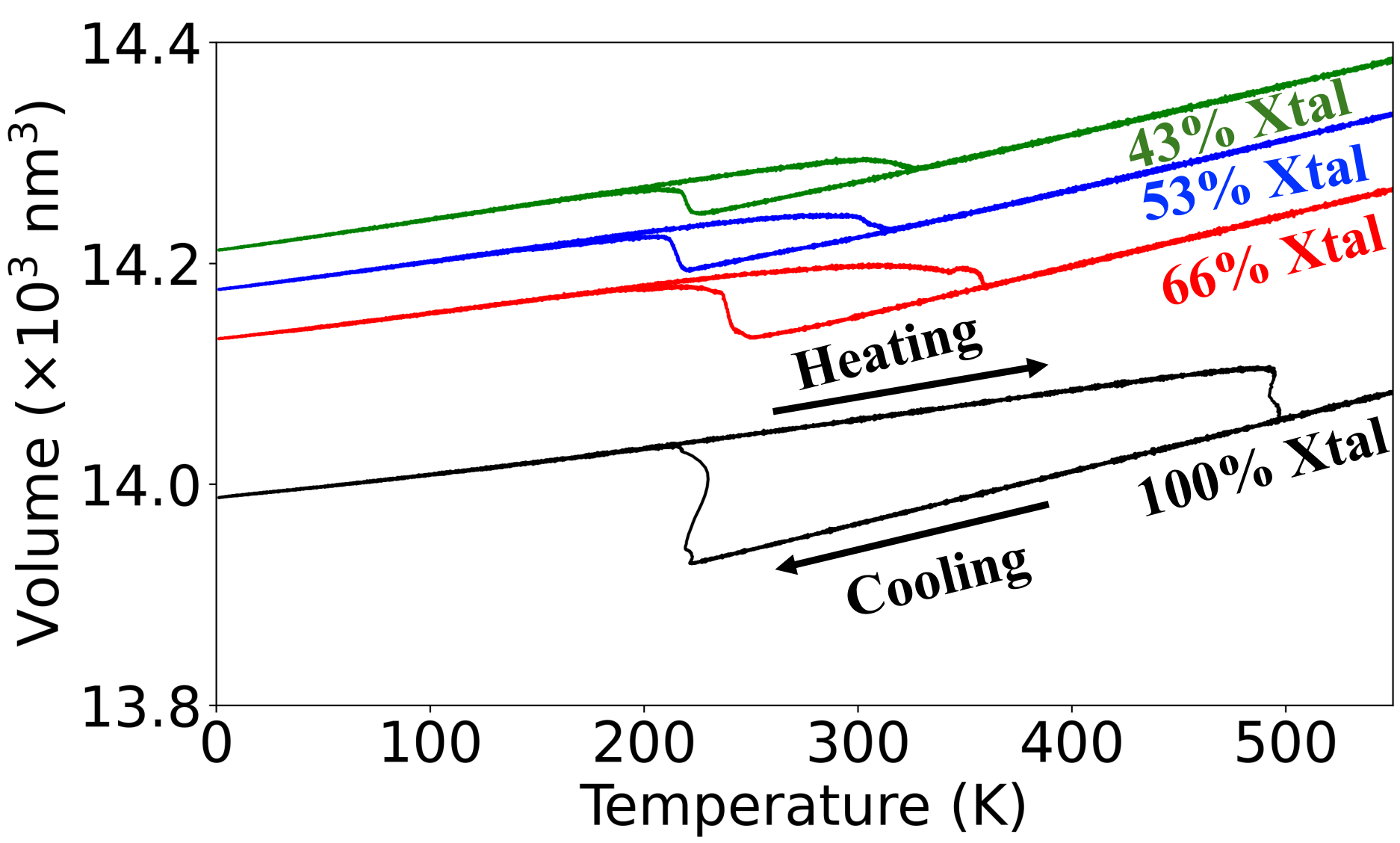

**SI Figure 2.** Evolution of the simulation cell volume as a function of temperature during cooling and heating for various CAS-NiTi systems. Green, blue, red, and black colors are used for 43%, 53%, 66%, and 100% Xtal CAS-NiTi.

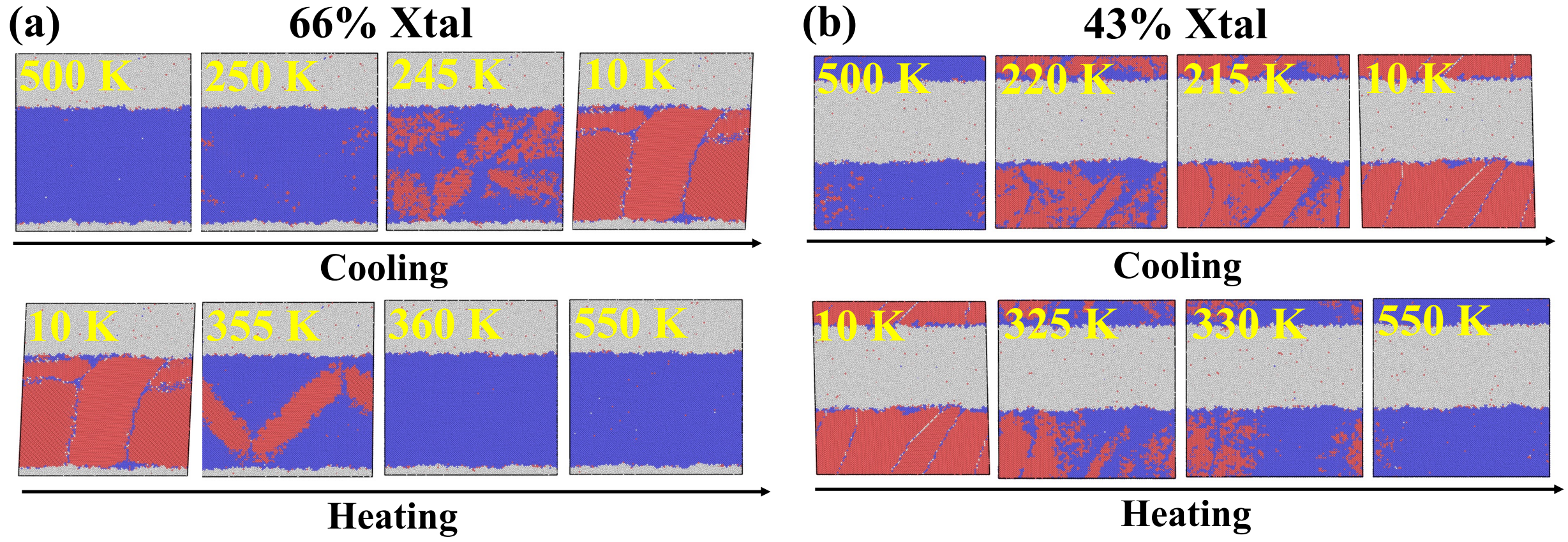

**SI Figure 3.** Atomic configurations during the forward (cooling) and reverse (heating) martensitic transformations for (a) 66% Xtal and (b) 43% Xtal CAS-NiTi systems. Snapshots correspond to selected temperatures during cooling (500 K, $M_s$, $M_s - 5$K, and 10 K) and heating (10 K, $A_f - 5$K, $A_f$, and 550 K). Atoms are colored according to the Polyhedral Template Matching (PTM) analysis: austenite (blue), martensite (red), and amorphous (gray).

## 2. Stress-induced martensitic transformation

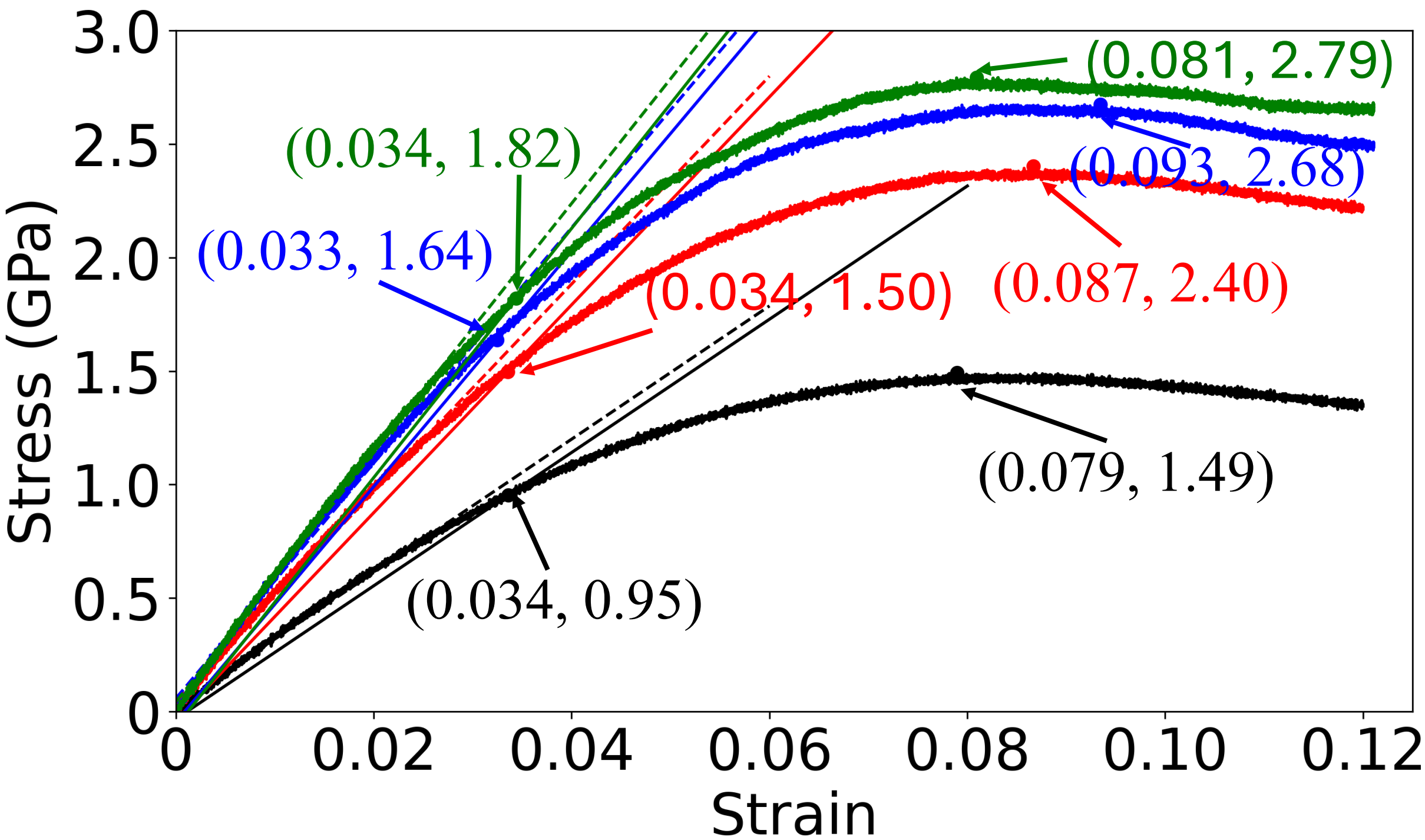

**SI Figure 4.** Representative loading stress–strain curves of the investigated CAS-NiTi systems showing the procedure used to determine the elastic modulus, the 0.2% offset critical stress for stress-induced martensitic transformation, and the maximum pre-plateau stress. The elastic modulus was obtained from a linear fit to the stress–strain response over the strain range of 0-0.03, while the 0.2% offset critical stress was determined using the corresponding offset line. The maximum pre-plateau stress was defined as the peak stress immediately preceding the transformation plateau.

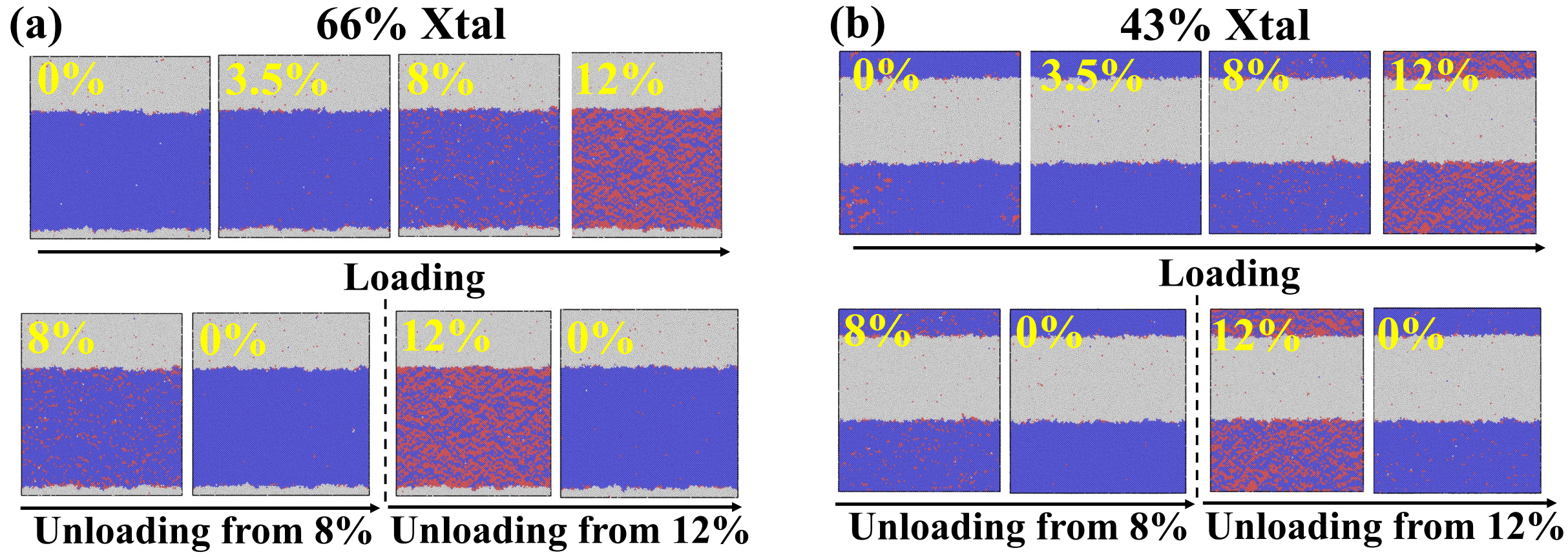

**SI Figure 5.** Atomic snapshots illustrating the evolution of stress-induced martensitic transformation during loading and unloading in (a) 66% Xtal and (b) 43% Xtal CAS-NiTi systems under uniaxial tensile loading along the [001] direction at 500 K. Snapshots during loading correspond to 0%, 3.5%, 8%, and 12% strain. Unloading is performed separately from maximum strains of 8% and 12%, and the corresponding unloaded configurations are shown to illustrate the reversibility of the transformation and the remnant martensite